\documentclass[aps,epsfig,float,twocolumn,pre,showpacs]{revtex4}

\newcommand{\be}{\begin{equation}}
\newcommand{\ee}{\end{equation}}

\usepackage{graphicx}

\begin{document}
\title{Transport on weighted Networks: when the correlations are
independent of the degree}

\author{Jos\'e J. Ramasco} \email{jramasco@isi.it}
\affiliation{CNLL, ISI Foundation, Viale S. Severo 65, I-10133 Torino, Italy}

\author{Bruno Gon\c calves} 
\affiliation{Physics Department, Emory University, Atlanta, Georgia 30322, USA}

\date{\today}

\begin{abstract}

Most real-world networks are weighted graphs with the weight of the edges 
reflecting the relative importance of the connections. In this work, we study
non degree dependent correlations between edge weights, generalizing thus the
correlations beyond the degree dependent case. We propose a
simple method to introduce weight-weight correlations in topologically 
uncorrelated graphs. This allows us to test different measures to 
discriminate between the different correlation types and to
quantify their intensity. We also discuss here the effect 
of weight correlations on the transport properties of the networks, showing 
that positive correlations dramatically improve transport. Finally, we give 
two examples of real-world networks (social and transport graphs) in which
weight-weight correlations are present.

\end{abstract}


\maketitle

\section{Introduction}

Complex networks have proved to be useful tools to explore natural 
or man-made phenomena as diverse as the Internet \cite{romu-book}, human 
societies \cite{newman-review}, transport patterns between 
airports \cite{alain04,roger05} or even metabolic reactions in the interior 
of cells \cite{almaas04}. The vertices in the networks represent the elements
of the system and the edges the interactions between them. The study of
the topology of the network provides valuable information on how
the basic components interact. While the existence or not of an edge is 
already informative, in many cases, as those listed above, the
interactions can appear on different levels. The bandwidth between 
two servers on the Internet for instance is not a flat quantity equal for all 
pairs, it depends on the importance of the servers as well as on the traffic 
expected. This fact led to the introduction of 
weighted graphs as a more accurate way to describe real networks 
\cite{yook01,newman01}.
Weighted graphs are complex networks where the edges have a magnitude 
associated, a {\it weight}. The weight accounts for the quality of 
a connection. The existence of a distribution of weights 
dramatically alters transport properties of networks like the
geometry of the optimal paths  \cite{braunstein03,goh05,wu06,chen06}, the 
spreading of diseases \cite{gang05} or the 
synchronabizability of 
oscillators \cite{zhou05}. Most previous studies have been carried out on
networks with uncorrelated weights on neighboring edges (those arriving at the
same node) even though most real cases possess correlation. Our aim here is to 
check how the presence of correlations can influence these results. 

There may be several kinds of correlations in random graphs \cite{romu06}. 
Recently, it has been shown that the edge weights in some 
real-world networks are related 
to other properties of the graph such as the degree (the number 
of connections a vertex has)
\cite{barrat04,macdonald05,barrat04b}. The weights were found to
follow, on average, a power law dependence on the degree. Several theoretical
mechanisms have been proposed to generate networks of this type 
\cite{models}. In this case, a clear
correlation is introduced between the weight of neighboring edges
but one may wonder whether this is the only possibility for weight correlations. 
If not, which other structures are possible? How can the correlations 
be quantitatively characterized? And most importantly, which influence 
do they have on the transport?

In this work, we address these questions. The organization of the manuscript is
the following. In the Section II, we present a model that allows to 
explore the different configurations 
for weight correlations independently of other properties of 
the network. Next, in the Section III, we consider and evaluate different 
magnitudes to estimate the type and intensity of weight-weight correlations. 
Section IV includes a study on how the presence of weight correlations  
affects transport. In Section V two examples of real-world
networks showing this type of correlations are discussed: the IMDB actor 
collaboration network and the traffic network between US airports. And finally,
we conclude and summarize in Section VI. 

\section{A simple method to introduce weight-weight correlations}

Let us start by defining a mathematical framework for the weight correlations.
From the point of view of an edge of weight $w$ with vertices with degree 
$k$ and $k'$ at its extremes, the joint probability that its neighboring edges
have a certain weight is 
given by 
\begin{equation}
P_{k \, k'} (w,w_1, \ldots ,w_{k-1}, w'_1, \ldots , w'_{k'-1}) .
\end{equation}
These functions contain all the information about both degree and 
weight distributions and correlations. However, a situation in which a full
hierarchy of such functions were needed to characterize the network would be 
hard to control from an analytical or numerical point of view. Therefore we 
will focus here only on the simplest scenario. In the same way 
 the Markovian condition 
 is a simplifying assumption for stochastic processes, we will consider 
 only correlations generated by two-point joint probability functions 
$P_{k \, k'}(w,w')$, and, among those, initially only the ones that are degree
independent given by functions of the type $P(w,w')$. 

\begin{figure}
\begin{center}
\leavevmode
\includegraphics[width=8.0cm]{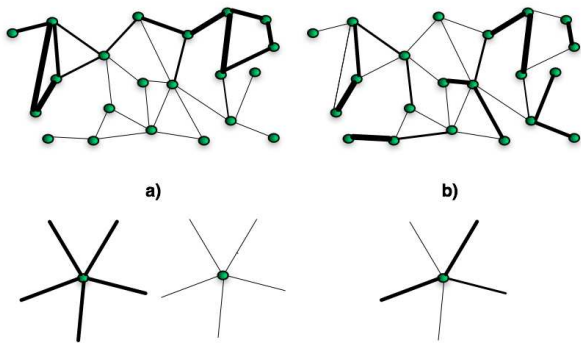}
\caption{(Color online) Two possible cases in networks with correlations in the 
link weight: a)
positively correlated nets and b) anticorrelated networks. The width of the line
of the links represents the value of the weight.}
\end{center}  
\end{figure}

In order to construct weighted networks along these lines, we use 
the so-called  Barab\'asi-Albert (BA) model \cite{barabasi99}, where new 
nodes entering the network connect to old ones with a probability proportional 
to their degree \cite{note1}. The networks generated by this model are
scale-free (their degree distribution goes as $P_k(k) \sim
k^{-3}$), have no degree-degree correlations, and their clustering coefficient 
(probability of finding triangles) tends to zero when the system size tends to
infinity. All this makes them ideal null models to test correlations between 
edge weights. Once the network is grown, a joint probability 
distribution for the link weights $P(w,w')$ and an algorithm for weight
assignation are needed. With the function $P(w,w')$ one 
can calculate the weight
distribution $P(w) = \int dw' P(w,w')$, and the conditional probability of
having a weight $w'$ provided that a neighboring link has a weight $w$, 
$P(w'|w) =  P(w,w')/P(w)$. We start by choosing
an edge at random and giving it a weight obtained from $P(w)$. Then we move to 
the nodes at its extremes and assign weights to the neighboring 
links. To do this, we follow a recursive method: if the edge from which the 
node is accessed has a
weight $w_0$, the rest, $w_1, \ldots, w_{k-1}$, are obtained from the
conditional distributions $P(w_i|w_{i-1})$. The recursion is necessary to 
increase the variability in case of anticorrelation (see below). If any of the 
links, $j$, has already a weight, it remains without change and its value 
affects 
the subsequent edges $j+1, \ldots k-1$. We repeat this process
until all the edges of the network have a weight assigned \cite{note3}.

For $P(w,w')$, 
we have considered different possibilities but here we will focus only on the
following three: 
\begin{equation}
\begin{array}{l}
P_+(w,w') = \frac{X_+}{(w+w')^{2+\alpha}} ,\\
\,\\
P_U (w,w') = \frac{X_U}{(w \, w')^{1+\alpha}} , \\
\,\\
P_-(w,w') =  \frac{X_-}{(w \, w' + 1)^{1+\alpha}} ,
\end{array}
\label{distr}
\end{equation} 
where $X_+ = 2^\alpha \alpha (1+\alpha)$, $X_U = \alpha^2$ and 
$X_- = \alpha^2/_2F_1(\alpha,\alpha,1+\alpha,-1)$ are the normalization factors
for the distributions on the domain of weights $(1,\infty)$, and 
$_2F_1()$ is Gauss hyper-geometric function \cite{book-hyp}. 
Without losing generality, we have chosen these particular functional forms due to 
their analytical and numerical tractability. The
distributions generated by Eqs. (\ref{distr}) asymptotically
decay as $P(w) \sim w^{-1-\alpha}$. The reason to use power-law decaying
distributions is that empirical networks 
commonly show very wide weight distributions
that in a first approach can be modeled as power-laws 
(see Fig. 6 and Refs. \cite{almaas04,alain04,roger05,ramasco06a}). 
We name the functions as $+$ (positively correlated),
$-$ (anticorrelated) and U (uncorrelated) because the average weight, 
$\langle w \rangle(w_0) = \int dw \, w \, P(w|w_0)$, 
obtained with the
conditional probabilities from a certain seed $w_0$ 
grows as $\langle w \rangle_+(w_0) = (1+\alpha+w_0)/\alpha$, decreases as 
$\langle w \rangle_-(w_0) = (\alpha+1/w_0)/(\alpha-1)$ and remains 
constant $\langle w \rangle_U = \alpha/(\alpha-1)$, respectively. This 
means that in $+$ networks the links of each node 
tend to be relatively uniform in the weights (see Fig. 1a), with separate 
areas of the graph concentrating the strong or the weak links, while in the 
negative case links with high and low weights are heavily mixed.

From a numerical point of view, we have checked how the variables 
to measure vary with the network size $N$. In the following, most 
results are shown for $N = 10^5$, which is big enough to avoid 
significant finite size effects. For each value of the
exponent $\alpha$ (from Eqs. (\ref{distr})) and for each type of correlations, 
we
have averaged over more than $600$ realizations. Note that we use $\alpha$ as a control 
parameter for the strength of the correlations. For
high values of $\alpha$, $P(w)$ decays very fast and the 
correlations become negligible, all links have almost the same weight. 
When $\alpha$ decreases however, the higher moments of $P(w)$ diverge and one 
would expect the correlations to be
more prominent.

\begin{figure}
\begin{center}
\leavevmode
\includegraphics[width=7.cm]{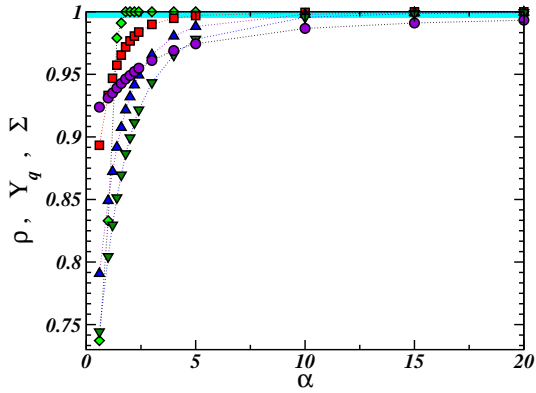}
\caption{(Color online) The rates $\Sigma_w$ (diamonds), $\Upsilon_q$, and
$\rho$ (circles)  for the
"+" distribution model as a function of $\alpha$. The highlighted area encloses 
the resolution limit of the estimators. The rate $\Upsilon_q$ is
shown for several values of $q$: $q = 2$ (squares), $q = 4$ (up triangles) and
$q = 6$ (down triangles).}
\end{center}  
\end{figure}

\section{Measures of weight correlations}

After a look at the sketch of Fig. 1, the first estimator to consider in order
to estimate weight correlations is the standard deviation of the weights of
the links arriving at each node. If the weights are relatively 
homogeneous, the standard 
deviation will be lower compared with its counterpart in
a randomized instance of the graph. The opposite will happen if the correlations
are negative as in the case b of Fig. 1. More specifically, for a  generic node 
of the network $i$, $\sigma_w(i)$ can be defined as
\begin{equation}
\sigma_w^2(i) = \sum_{j \in \nu(i)} (w_{ij} - \langle w\rangle_i)^2 ,
\end{equation}
where $\nu(i)$ is the set of neighbors of $i$ and $\langle w\rangle_i$ is the 
mean value of the weight of the links arriving at $i$. Once the 
deviation 
is calculated for each node, an average can be taken over the full network 
getting $\langle \sigma_w \rangle = (1/N) \sum_i \sigma_w(i)$. Then to
evaluate the effects of weight correlations, it is necessary to compare the 
value of $\langle \sigma_w \rangle_{org}$ obtained for the original network 
with that measured on uncorrelated graphs. It is, of course, important that the
statistical properties of such uncorrelated graphs are as close as possible to
those of the original graph. The 
most accurate procedure consists in disordering only the weights of the links of 
the original network. To do so, we interchange the weight of each link with
that of a randomly selected edge preserving so the 
weight distribution 
$P(w)$ and the network 
topology; $i.e.$, degree distribution, degree-degree correlation, clustering, 
etc, remain invariant. Once $\langle \sigma_w \rangle$ is
estimated for the original graph and for an ensemble of
weight-reshuffled instances of it, the rate 
\begin{equation}
\Sigma_w = \frac{\langle \sigma_w \rangle_{orig}}{\langle \sigma_w \rangle_{rand}} 
\end{equation}
can be calculated. If $\Sigma_w > 0$, the weight correlations in the original 
graph will be as in
the case b of Fig 1. If it is identically one, there will be no weight correlations
and if  $\Sigma_w < 0$ the correlations will be as in Fig. 1a. The behavior of 
$\Sigma_w$ for the positive and negative models proposed in the previous section
is displayed in Fig. 2. The first thing to remark is that indeed $\Sigma_w$
can distinguish between the three cases. Moreover, it provides a first method 
to quantitatively estimate of the intensity of the weight correlations.

\begin{figure}
\begin{center}
\leavevmode
\includegraphics[width=7.cm]{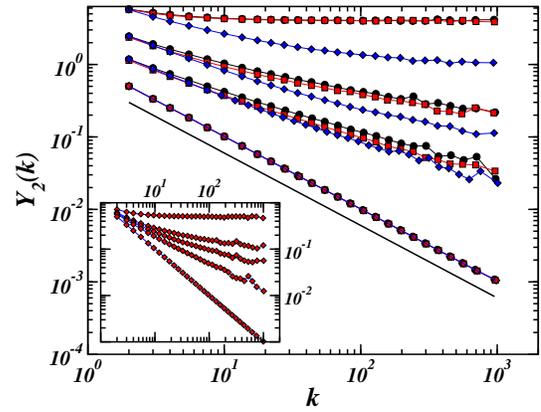}
\caption{(Color online) Plot of the disparity as a function of the degree, 
the triplets represent the graphs generated using the "+" (blue
diamonds), the uncorrelated  (red squares) and "-"  (black circles)
distributions. Each triplet correspond to a different value of $\alpha$, from
bottom to top $\alpha = 20, 1.5,1.2$ and $0.5$. The straight line has slope $-1$
and is meant as a guide to the eye. The triplets have been shifted
upwards to facilitate visibility. The inset shows the disparity
versus $k$ for the $U$ case and for the values of $\alpha$ from bottom to
top $\alpha = 20, 2, 1.5, 1.2, 1$ and $0.5$.}
\end{center}  
\end{figure}

A similar result can be obtained with a magnitude that was previously
discussed in the literature \cite{marc03,almaas04}. Its name is {\it disparity} 
and was introduced in the context of weighted graphs by 
Barth\'elemy {\it et al.} as a way to estimate how homogeneous the weights of 
the links arriving at a vertex are. The generalized disparity of node 
$i$, $Y_q(i)$, is defined as
\begin{equation}
Y_q(i) = \frac{\sum_{j \in \nu(i)} w_{ij}^q}{s_i^q} ,
\label{dispa}
\end{equation}
where $s_i$ is the strength of $i$, $s_i = \sum_{j \in \nu(i)} w_{ij}$.
If all the links of a node have a
similar weight, their value will be $w \approx s/k$, and therefore the 
disparity decays as $Y_2(k) \sim 1/k^{q-1}$. On the other hand, if the vertex 
strength is essentially due to the weight of a single link, $Y_q(k)$ 
will tend to a constant. Typically instead of a generalized $Y_q$, most of the
works in the literature has focused on $Y_2$, for which is commonly 
reserve the name of disparity. This latter magnitude can be related 
to $\sigma_w$ by the following expression for each node $i$ of the network 
\begin{equation}
Y_2(i) = \frac{1}{k_i} \, \left( 1 + \frac{\sigma_w^2(i)}{\langle w\rangle_i^2}
\right) ,
\end{equation} 
where $\langle w\rangle_i$ is the average weight of the links of $i$ and $k_i$
its degree. An important question to mention here is that the profiles of 
$Y_q(k)$ depend on the weight distribution, even for completely uncorrelated
graphs. $Y_q(i)$ provides information on how different the weights of the 
links of
$i$ are but not on whether that particular configuration is or not a product of
randomness or correlations. The variation of $Y_2(k)$ with the exponent of the 
weight distribution $\alpha$ for the uncorrelated $U$-model can be seen 
in the inset of Figure 3. In the same Figure, we also show the behavior of 
$Y_2(k)$ for 
the other two models,  "$+$" and "$-$", for a few values of the exponent $\alpha$. 
As before, in 
order to estimate the importance of the weight correlations, the disparity of 
the correlated graph has to be compared with that obtained
from uncorrelated networks. To express this comparison in a single rate, we 
can take the average of the disparity over all nodes of the
network, $\langle Y_q \rangle$, and calculate
\begin{equation}
\Upsilon_q = \frac{\langle Y_q \rangle_{orig}}{\langle Y_q \rangle_{rand}} .
\end{equation} 
The value of $\Upsilon_q$ for the positive model as a function of
$\alpha$ is displayed in Fig. 2. This magnitude is also able to
discriminate between the different correlations.

\begin{figure}
\begin{center}
\leavevmode
\includegraphics[width=7.cm]{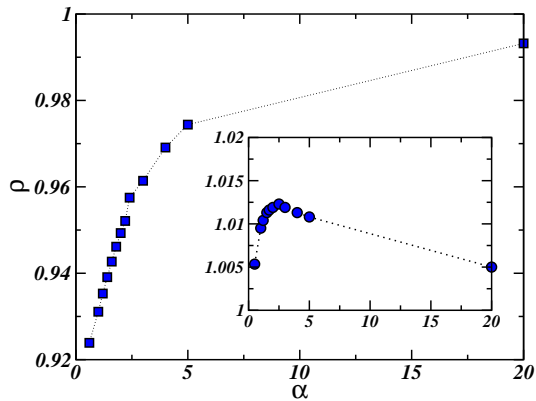}
\caption{(Color online) In the main plot, the dependence of $\rho$ on the 
exponent $\alpha$ for  
the positively correlated graphs generated with the "+" distribution of Eq. (\ref{distr}). In the
inset, the same function but for the graphs of the "-" distribution.}
\end{center}  
\end{figure} 

However, both $\Sigma$ and $\Upsilon_q$ have resolution problems. As can be seen
in Figure 2 for the positive model, if an area enclosing the numeric error 
is set immediately below one, the estimators $\Sigma$, $\Upsilon_q$
fall in turn relatively fast in that zone. The weights of the links in the "+" model 
are continuous variables
and therefore they are always correlated. Although, as explained
before, for higher values of $\alpha$ the effects of weight correlations can 
be weaker but still until $\alpha$ is not infinite they are not zero.
An ideal estimator should be able to distinguish the $+$ model from a
complete random configuration at very high values of $\alpha$. In this context,
$\Sigma_w$
seems to be the worst estimator. $\Upsilon_q$ is better than $\Sigma_w$ and 
improves 
the higher $q$ becomes. 
The
reason for this behavior is that  these
magnitudes are not only estimating how wide the spectrum of values of
$w$ for a node is, they also supply information on the shape of the 
distribution of those values. As an 
example, let us consider a node with $k$ links. The value of $\sigma_w$ is
higher if $k-1$ of them have weight $a$ and the remaining weight $b$,
$\sigma_w = |a-b| \, \sqrt{k-1}/\sqrt{k}$, than if the distribution is more 
symmetric, let us say, with half
of them with $w = a$ and the other half with weight $b$, 
$\sigma_w = |a-b|/(2 \sqrt{2})$. The goal here is to study how different the
amplitude of the weight values is compared with a
random configuration of weights, hence the extra information contained in 
$\sigma_w$ 
or $Y_q$ can be neglected. An ideal estimator for weight correlations only
needs to consider the interval $|a-b|$. Following this idea, 
we define  the 
{\it range} for a node $i$ as    
\begin{equation}
r_i = \frac{w_{max}(i)-w_{min}(i)}{w_{max}(i)+w_{min}(i)} ,
\end{equation}
where $w_{max}(i)$ and 
$w_{min}(i)$ are respectively the maximum and minimum weights of the edges 
of $i$. The denominator is a normalization factor to keep $r_i$ 
between zero and one. Note that $r$ has a similar behavior to $Y_q$ in the limit
 $q \to \infty$: if all the weights are equal $r = 0$, and also $Y_q \sim k^{1-q} \to 0$ 
 if $q \to \infty$. On the other hand, if the weight of link dominates 
 the others $r \to
1$, $Y_q \to 1$ too if $q \to \infty$. As before, to generate a correlations
estimator, the average 
of $r_i$, $\langle r \rangle$, 
can be taken over
all the nodes of the network and contrasted with the equivalent value obtained 
from
a set of weight-reshuffled instances. We will call $\rho$ to the rate between
these two quantities,  
\begin{equation}
\rho = \langle r \rangle_{orig}/\langle r \rangle_{rand} .
\label{rho}
\end{equation}
If $\rho < 1$, the network displays positive weight 
correlations. The
stronger they are, the smaller $\rho$ becomes. Otherwise, if  $\rho > 1$, the 
weights are anticorrelated. 
$\rho = \Upsilon = \Sigma = 1$ is the limit of uncorrelated networks. As can be
seen in Fig. 2, $\rho$ is a much more acute estimator of weight 
correlations than $\Upsilon_q$ or $\Sigma$. Hence, from
now on we will present our results as function of it. The variation 
of $\rho$ with $\alpha$
is displayed in Fig. 4 for the "$+$" and "$-$" models. The intensity of the 
correlations for the "$+$" model
grows when $\alpha \to 0$  ($\rho_+$ decreases for smaller $\alpha$), while for the
negative case $\rho_-$ initially
grows, peaks around $\alpha \approx 2.5$ and then tends to one again for 
smaller $\alpha$. 

Another question that we are now in position to board is in which 
way a relation between weight and degree affects the weight-weight 
correlations. As mentioned in the introduction, networks of this kind
 may display quite different transport properties from their unweighted 
counterparts \cite{macdonald05,luca,gang05,wu06}. Usually, the weights of
these networks are obtained by means of a relation of the kind 
$w_{ij} \sim (k_i \, k_j)^\delta$ \cite{barrat04}. The result is that 
provided that the degree is an 
"{\it a-priori}" characteristic 
that equally influence all the edges of a vertex, the weight of the links is 
positively correlated. The networks created in this way show correlations 
similar to our "$+$" model (regardless of the sign of the exponent $\delta$)
\cite{note2}. For instance, for $\delta = \pm 0.5$ the value of $\rho$ is 
in both cases $\rho \approx 0.832$.

\section{Transport properties}

Let us focus now on the transport and how it varies with the presence of
weight-weight correlations. Several measures have been proposed to study
transport \cite{wu06,transport}. In this work we will concentrate on the size
and weight of the "{\it superhighways}" as recently introduced by Wu 
 {\it et al.} \cite{wu06}. If the edges of an uncorrelated network
are severed following an increasing order from small to higher weights, the 
percolation threshold is eventually reached. The remaining connected graph after 
the process is over, the incipient percolation cluster (IIC) or superhighways, 
holds most of the traffic of the original network. Regarding this procedure,
there are several questions to mention. First, along this work we have assumed
that higher weights imply better transport properties. This is also the case for
the two empirical examples discussed below; the weights represent number of
passengers in one network and number of collaborations done together in the 
other. However, for some other graphs, the weights can mean higher resistance 
to transport. And, therefore, the superhighways must be obtained with the same
procedure but cutting the links with highest weights first. In such
circumstances, the lower the
total weight of the superhighway is, the better the transport results. 

\begin{figure}
\begin{center}
\leavevmode
\includegraphics[width=7.cm]{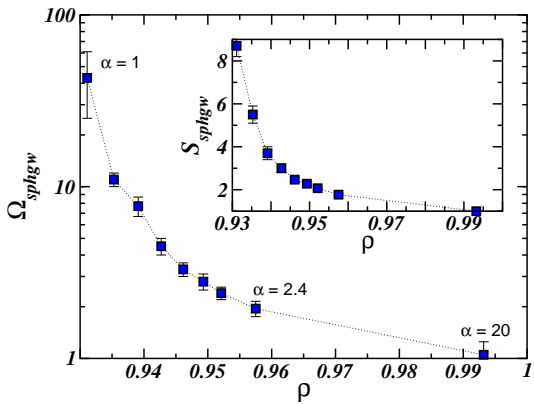}
\caption{(Color online) Relation between the intensity of the correlations, 
estimated with $\rho$, and the 
change in the weight of the superhighways for the
positively correlated networks generated with the "+" model of Eq. (\ref{distr}). The values
of $\alpha$ go from $\alpha =1$ for the first point on the left to $\alpha = 2.4$ 
in middle changing regularly by $0.2$ for each point. The last point of the
right hand side is for 
$\alpha = 20$.}
\end{center}  
\end{figure} 

Another question is that the percolation threshold depends on the topological
characteristics of the network. For uncorrelated undirected graphs, it is
attained when in the process of severing random links, the rate $\langle k^2 \rangle/ \langle k
\rangle$ of the remaining graph approaches two, $\langle k^2 \rangle/ \langle k
\rangle = 2$ \cite{transport}. For directed 
networks 
or networks with {\it high clustering}, the threshold does obey 
different expressions \cite{marian05}. In the next section, we will further 
discuss this point as well as the numerical determination of the
percolation threshold for our two empirical networks. 

Finally, it is worthy noting that this 
method to estimate superhighways cannot
be applied directly to networks with weight correlations. If the high weight
links are concentrated in some areas of the graph, cutting the weak links is 
not going to affect those areas. This can change the percolation 
criterion, i.e. $\langle k^2 \rangle/ \langle k \rangle = 2$ cannot be used, 
although the percolation threshold that depends solely on the 
topology remains the same as in an uncorrelated graph. 

The goal in our case is to compare graphs with and without weight correlations and to
quantitatively estimate the effects of these correlations. 
The method used in practice is to disorder the weights of 
the links of each correlated network. Then we estimate the superhighways
of the randomized graphs and measure how many edges must be cut on 
average to attain the percolation threshold. Reaching the
percolation may have some numerical problems \cite{wu07} so the process 
of reshuffling the weights must be repeated several 
times. Next, we cut the 
same number of 
links in the correlated network (again going 
from lower to higher values of $w$) and compare the size and the weight
of the biggest remaining connected cluster ($W_{sphgw}$) with the average of 
those found for the randomized graphs. In this way, we obtain
\begin{equation}
\begin{array}{l}
\Omega_{sphgw} = \frac{W_{sphgw}(orig)}{\langle W_{sphgw}(rand) \rangle}  ,\\
\, \\
S_{sphgw} = \frac{\mbox{Number nodes in superhighway (orig)}}{\langle
\mbox{Number nodes in superhighway (rand)}\rangle} .
\end{array} 
\end{equation}
The 
results, displayed in
Fig. 5, show that, in general, positive correlations play a decisive role on
the value of $W_{sphgw}$, increasing it by orders of magnitude. The smaller is
$\rho$ respect to the unit, the stronger the effect of the weight correlations
on the transport becomes.  
This phenomenon may be understood by keeping the
analogy with the roads: the transport improves if the 
highways are connected together forming a communication 
backbone as large as possible. Anticorrelated networks, on the other hand, exhibit smaller
superhighways than their randomized counterparts although the effect is subtle. 
  
So far we have discussed models for which the
weight correlations are independent of other structural factors. In general, 
there may be other aspects influencing the transport 
properties of a graph. If several compete, as it happens in the case
$w_{ij} = (k_i \, k_j)^{\delta}$, with $\delta < 0$, between the degree and the weight of the
connections of a node, the transport capability of the network may suffer. For example, we 
measure $\Omega_{sphgw} = 75(1)$ for networks of size $N = 10^5$ and $\delta = 1/2$, while  
$\Omega_{sphgw} = 0.014(1)$ if $\delta = -1/2$.

\begin{figure}
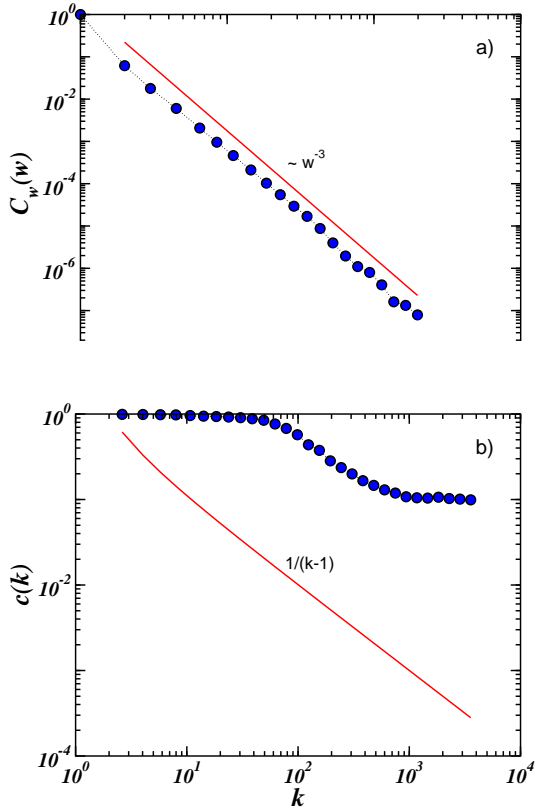

\begin{center}
\leavevmode
\includegraphics[width=7.cm]{fig6a.eps}
\qquad
\includegraphics[width=7.cm]{fig6b.eps}
\caption{(Color online) In a), cumulative weight distribution $C_w =
\int_w^\infty dw\, P(w)$ for IMDB actor collaboration network. In the bottom, in
b), the clustering as a function of the degree. The line below corresponds to
the function $1/(k-1)$.}
\end{center}  
\end{figure}

\section{Empirical networks}

Finally, we also consider a couple of real-world examples. First 
the IMDB movie database with $383640$ actors that are
connected together whenever they have shared
a common movie \cite{barabasi99,actors}. This network is formed by the union 
of cliques, which
means that the  number of links is high, $15\, 038 \, 083$ in total. The 
weight of each 
link represents the number of 
times a partnerships has been repeated. Higher values
of the weight imply an increased probability of information transfer between 
two individuals. The cumulative distribution 
of weights for this network can be seen in Figure 6a. It shows a very wide
functional form that can be well represented by a power-law. The presence of 
weight correlations in collaboration 
networks have
been discussed using a different technique in Ref. \cite{ramasco07}. Here we will
focus on the results obtained with $\rho$, $\Omega_{sphgw}$ and $S_{sphgw}$. 

First of all, it is important to note that collaboration networks typically do
not show a relation between the weight of the links defined in this way, or as
{\it social closeness} \cite{newman01}, and the degree of the nodes 
\cite{ramasco07}. Hence the weight
correlations, if exist, are not a product of this type of relation. And, indeed,
they exist since the actor network presents a value of $\rho = 0.268(1)$. 

\begin{figure}
\begin{center}
\leavevmode
\includegraphics[width=7.cm]{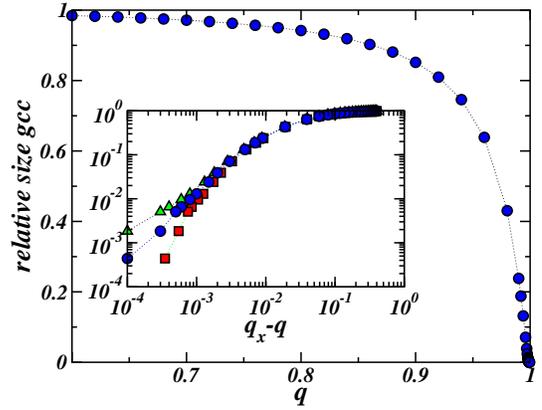}
\caption{(Color online) Relative size of the giant component as a function of
the removal probability $q$ for the actor collaboration networks. In the inset,
the same quantity but in log-log and versus $q_*-q$. Three values of $q_*$ are
displayed: $q_* = 0.9987$ (triangles up), $q_c = 0.9990$ (circles) and $q_* =
0.9993$ (triangles down).}
\end{center}  
\end{figure}

The measure of the superhighways in this case poses a certain level of 
challenge. The actor
collaboration network, as many social networks, presents high clustering.
The clustering of a node $i$ is defined as 
\begin{equation}
c_i = \frac{2 s_i}{k_i (k_i-1)} ,
\end{equation}
where $s_i$ is the number of connections between the neighbors of $i$ and $k_i$
stands for its degree. The average over all the nodes of the
network with the same degree $k$ can be then taken to obtain $c(k)$, which is 
depicted in Figure 6b for this network. In the same plot, it
is also included the curve $1/(k-1)$. The comparison is 
necessary because it has been recently shown that the percolation threshold of
a graph is highly dependent on the clustering \cite{marian05}. In fact,  
clustered networks can be classified into two major groups: those
with {\it weak} and those with {\it strong} clustering. The
difference between the two groups is whether $c(k)$ decays as $1/(k-1)$ (weak
clustering) or in a slower way (strong). The actor network clearly falls into this latter
group. 

For weak clustered networks, it is possible to find a generalization of the
percolation threshold condition mentioned in the previous section ($\langle k^2
\rangle/\langle k \rangle = 2$) \cite{marian05}. However, as far as we know, 
nothing similar has been proposed for strong clustered graphs. Therefore, we
will be force to use a more pedestrian technique to estimate the percolation
threshold of the actor collaboration network. In Figure 7, the behavior of the
rate between the size of the giant component $gcc$ and its original value
$gcc_0$ is displayed as a function of the percentage of
links severed $q$. A continuous transition can be observed with this rate as
order parameter. Assuming functional form of the type $gcc/gcc_0 \sim
(q_c-q)^\beta$, we find that the critical point happens at a remotion
rate of $q_c \approx 0.9990(3)$ (see the Inset of Fig. 7). The point in which 
the condition $\langle k^2 \rangle/\langle k \rangle = 2$ is fulfilled, for 
instance, lies in a smaller value $q = 0.9976(1)$. Once $q_c$ has been measured,
we can proceed as in the previous section, cutting a fraction
$q_c$ of links following an ordered sequence from lower to higher values of the
weight and comparing the results with those obtained for a graph in which the
weights of the links have been reshuffled. The rates for the total weight 
of the superhighways for a few values of $q$ are
$\Omega_{sphgw}(q=0.9976) = 3.2(1)$, $\Omega_{sphgw}(q=0.9987) = 17(5)$, and       
 $\Omega_{sphgw}(q_c=0.9990) = 106(30)$. As can be seen, finding the value of 
 $\Omega_{sphgw}$ requires a fine determination of $q_c$. Even so, the high
 values of this rate gives us a clear feeling of the importance that the weight
 correlations have on the transport properties of these real world graphs. The
 values of $S_{sphgw}$ that we find for the same removal rates $q$ are 
 $S_{sphgw}(q=0.9976) = 0.62(2)$, $S_{sphgw}(q=0.9987) = 3.2(1.5)$, and       
 $S_{sphgw}(q_c=0.9990) = 20(7)$, respectively.

\begin{figure}
\begin{center}
\leavevmode
\includegraphics[width=7.cm]{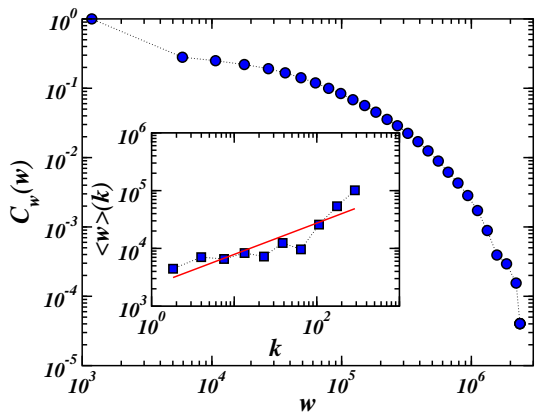}
\caption{(Color online) In the main plot, the cumulative weight distribution for
the US airport traffic network. The average weight of the outgoing connections
is displayed in the inset as a function of the out-degree.}
\end{center}  
\end{figure}

The second example is a 
network composed by $1278$ US airports. A directed edge connects two airports
whenever there is a direct flight between them. The weight of the links 
represent in this case the number of 
passengers on that traject 
during $2005$ \cite{airports}. The cumulative weight distribution of this
network is displayed in Figure 8. The weight distribution is in this case also
wide, although clearly it does not follow a power-law decay. As can be seen 
in the Inset of that Figure, the  average weight of the 
outgoing links exhibits a dependence on the out-degree of the nodes, 
$\langle w_{ij} \rangle \sim (k_i k_j)^{0.42}$. Therefore it is not strange 
that the value of $\rho$ that we
measure, $\rho = 0.983(1)$, delates the presence of 
positive
weight correlations. Since 
the number of passengers in each direction can be different, to calculate the 
superhighways it is necessary to
generalize the concept to directed graphs. This means to study 
the incipient 
strongly connected component (SCC) instead of the incipient percolation 
cluster.  Applying the same technique as the one illustrated in Fig. 7, 
we get a value for the
critical removal of $q_c = 0.988(2)$. The corresponding 
rate $\Omega_{sphgw}$ is $\Omega_{sphgw} = 2.61(7)$.

\section{Conclusions}

In summary, we have explored how correlations between neighboring edge weights 
can occur in random networks. The high (low) weights can appear concentrated 
in certain areas of the graph, a configuration that has a considerable effect 
on transport properties. To study this phenomenon, we have proposed a simple 
method to 
introduce weight correlations in otherwise uncorrelated graphs. These
models show that weight correlations can appear independently of any other
property of the network, although they could be also coupled to some
characteristic of the vertices such as the degree, hidden variables, etc. This 
method allow us to study,
not only qualitatively but also quantitatively, the type and intensity of these 
correlations. Leading us to test several estimators: 
$\sigma_w$, the generalized disparity and the range $\rho$, being the latter 
the best of the three.

Once we found a tool to measure the intensity of weight correlations, we have
focused on how the transport properties of the network become affected by these
correlations. The so called superhighways of our model "$+$" have been studied 
as
a function of the intensity of weight correlations. The conclusion, that seems
to be generalizable to other networks, is that stronger 
(positive) correlations imply bigger and weightier superhighways, improving thus
the performance of the network on transport in orders magnitude.

Finally, we 
have also considered data from two real-world networks, a
collaboration graph and a transportation (airports) network. Both cases present
positive weight-weight correlations. The results on their superhighways, 
$\Omega_{sphgw}$, also
prove that weight correlations are without doubt an important factor to take 
into account in the 
study of transport on real networks.

{\it Acknowledgments---}  The authors thank Stefan Boettcher and Eduardo 
L\'opez for useful discussion and comments. Funding from the NSF under 
grant 0312510 and from Progetto Lagrange was received.


\begin{thebibliography}{99}
\bibitem{note1} We also tested other growth models but the results do not 
change significantly.
\bibitem{note3} After the assignation process is over, we have checked that the weights are
distributed as $P(w)$ for every case.
\bibitem{note2} This is not necessarily true for dynamical models 
\cite{models} where new edges are introduced at the same time
that the weight of the old ones is increased.
\end{thebibliography}
\end{document}